# An Integrated Framework of Decision Making and Motion Planning for Autonomous Vehicles Considering Social Behaviors

Peng Hang, Chen Lv, Senior *Member, IEEE*, Chao Huang, Jiacheng Cai, Zhongxu Hu, and Yang Xing

*Abstract*—This paper presents a novel integrated approach to deal with the decision making and motion planning for lane-change maneuvers of autonomous vehicle (AV) considering social behaviors of surrounding traffic occupants. Reflected by driving styles and intentions of surrounding vehicles, the social behaviors are taken into consideration during the modelling process. Then, the Stackelberg Game theory is applied to solve the decision-making, which is formulated as a non-cooperative game problem. Besides, potential field is adopted in the motion planning model, which uses different potential functions to describe surrounding vehicles with different behaviors and road constrains. Then, Model Predictive Control (MPC) is utilized to predict the state and trajectory of the autonomous vehicle. Finally, the decision-making and motion planning is then integrated into a constrained multi-objective optimization problem. Three testing scenarios considering different social behaviors of surrounding vehicles are carried out to validate the performance of the proposed approach. Testing results show that the integrated approach is able to address different social interactions with other traffic participants, and make proper and safe decisions and planning for autonomous vehicles, demonstrating its feasibility and effectiveness.

## I. Introduction

AUTONOMOUS vehicles mainly consist of four functionality modules, namely the environment perception, decision making, motion planning, and motion control [1]. Among the above four modules, decision-making and motion planning, which bridge the environment perception and motion control, are regarded as the AV's brain and of great importance [2]. Positioning next to each other, the modules of decision making and motion planning are highly correlated in terms of the functionality and resultant performance. Thus, the design of decision making for AV should take the feasibility of motion planning into consideration, and on the other side, the motion planning module should be also developed based on the out behavior of decision making.

In recent years, many researchers have conducted comprehensive studies on decision making and motion planning of AVs.

Markov Decision Process (MDP) approach and Bayesian networks are widely used for threat assessment to assist decision making [3]-[5]. To make AVs move safely and efficiently amid pedestrians, partially observable MDP is studied for robust decision making under uncertainties [6]. However, MDP approach has a high computational complexity. Threat measurement, Bayesian networks and time window filtering are combined together to provide an optimal decision making for AVs [7]. Based on the Multiple Attribute Decision Making method, a maneuver decision-making algorithm is designed to deal with complex urban environment [8]. Besides, Multiple Criteria Decision Making method is also widely used to select the most appropriate driving maneuver [9].

In addition to the aforementioned decision-making methods, data-driven learning-based decision-making methods, including Support Vector Machine (SVM) [10], Clustered SVM [11], Extreme learning machine (ELM) [12], Kernel-based ELM [13], reinforcement learning (RL) [14], Deep Neural Networks (DNN) [15], are gaining popular. Since RL can provide many benefits in solving complexed uncertain sequential decision problems, the authors in [16] develop a decision-making system by integrating MDP and RL. A human-like decision-making system is built based on DNN, which can adapt to real-life road conditions [17].

Although, there have been lots of progress made, decision making of AVs considering social interactions of vehicles has been rarely reported. To further advance the decision making of AVs, capturing the features during vehicle interactions is of great significance, and game theory is an effective method to address this problem. Based on game theory, a human-like decision-making system is proposed for autonomous vehicles to realize automatic lane change and car following [18], [19]. In a connected vehicular environment, game theory is used to predict the lane change behavior [20]. Moreover, Stackelberg Game theory is adopted to solve the merging problem of AVs in [21]. Considering the interactive behaviors and characteristics between human-driven vehicles, the designed decision-making system is expected to be human-like.

In terms of motion planning, it includes both longitudinal

This work was supported in part by the SUG-NAP Grant (No. M4082268.050) of Nanyang Technological University, Singapore, and A*STAR Grant (No. 1922500046), Singapore.

P. Hang, C. Lv, C. Huang, J. Cai, Z Hu and Y. Xing are with the School of Mechanical and Aerospace Engineering, Nanyang Technological University, Singapore 639798. (e-mail: {peng.hang, lyuchen, chao.huang, caij0018, zhongxu.hu, xing.yang}@ntu.edu.sg)

(Corresponding author: Chen Lv)

and lateral planning. Longitudinal planning usually refers to velocity or acceleration planning for AV in the longitudinal direction. The algorithm of longitudinal planning is already matured, and it has been applied in existing Adaptive Cruise Control (ACC) systems, Cooperative ACC, and Autonomous Emergency Braking systems [22]-[24]. The lateral planning often refers to path planning, which is the current key focus of motion planning algorithm. Many algorithms have been studied for path planning, which can mainly be divided into five categories: the graph search, sampling algorithm, potential field, interpolating curve, and intelligent optimization [2].

Graph search algorithm is the most common seen and practical one for path planning. It includes A* algorithm [25], D* algorithm [26], Dijkstra algorithm [27], etc. Graph search algorithm has good performance in terms of collision avoidance. However, its optimal path is usually affected by the density of grid. Rapidly-exploring random trees (RRT) is a typical sampling algorithm, which is able to efficiently search the optimal path considering non-holonomic constraints [28]. However, the stability of RRT and the quality of its planned path need to be further improved. Potential field, which is based on the theory of force field, is another effective approach for path-planning [29]-[31]. It is usually combined with intelligent optimization methods to search the optimal path. The method of interpolating curves is the simplest and most widely used one. It typically includes line, circle, polynomial curve, spline curve, etc [32]. These curves are usually used in special scenarios.

To further advance the performance of AVs, in this study the decision making and motion planning modules are considered in an integrated manner. Moreover, the social behaviors, which can be reflected by driving styles and intentions of surrounding vehicles, are also considered. The decision-making process of an AV is formulated as a non-cooperative game, and the Stackelberg Game theory is applied to solve this problem. In addition, the potential field model is built during the development of the motion planning algorithm. Finally, the decision making and motion planning modules are integrated via the common feature identifications of social behaviors.

The rest of this paper is organized as follows. In Section II, problem formulation and high-level system architecture are presented. Then, non-cooperative game based decision-making framework is established in Section III. In Section IV, motion planning algorithm is developed using MPC. Moreover, the characterizations of social behaviors of surrounding traffic occupants are analyzed in Section V. Testing, validation and discussion of the proposed approach are given in Section VI. Finally, Section VII concludes the paper.

## II. Problem Formulation

As introduced in Section I, most of the current works investigate decision-making and motion planning separately. It poses an obvious drawback that the feasibility of motion planning is not fully considered when designing the decision making module. As a result, undesirable paths may be planned with occurrences of frequent braking and sharp steering maneuvers. In the process of modeling, the boundaries for the design of motion planner are usually taken as the constraints of decision making. Sequentially, after the decision made, motion planning will be conducted within the pre-defined constrains. If the constrains are set too narrow, feasible solutions may be hardly found by the motion planner. However, if the boundaries are set too large, as a result, the exploration space would be small, and the performance potential may not be fully exploited. To overcome the above drawback of the existing approach, in this work, decision making and path planning will be further investigated in an integrated manner.

Additionally, different drivers have distinguished driving styles, i.e. different drivers may make different decisions under a same condition. For instance, facing the overtaking behavior of an adjacent vehicle, aggressive drivers may choose to accelerate and stop the adjacent vehicle from overtaking. However, timid drivers may slightly decelerate and leave more space for others to pass through. Hence, ideally, the integrated decision making and path planning for AVs should be able to adapt to various social interaction behaviors. In this paper, three different driving styles of obstacle vehicles, i.e., aggressive, cautious and normal, will be considered in the proposed approach. We assume that the aggressive drivers care more about vehicle dynamic performance and travel efficiency, which means they would operate the steering wheel and/or press pedals more frequently. In contrast, the cautious drivers regard driving safety and ride comfort as top priorities. Thus, their operational actions are usually careful. The normal drivers, positioned in between, are more likely to do a trade-off between driving safety, travel efficiency, and ride comfort during their decision making procedures [33].

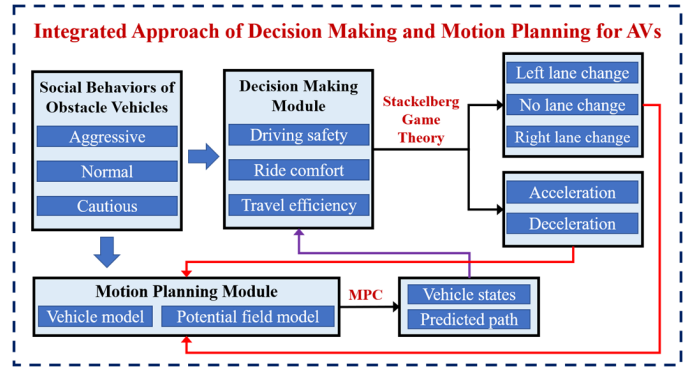

Fig. 1. Schematic diagram of the integrated framework for decision making and planning considering social behaviors.

There are many driving scenarios for algorithm development of AVs. Since lane-change is one of the most typical action in high way scenarios, it is taken as the target scenario for system design and proof of concept in this paper. The schematic diagram of the integrated approach of decision making and path planning for lane change of AV is displayed in Fig. 1. There exist three choices for lateral actions and two options for longitudinal motion. Considering the driving styles of obstacle vehicle, decision-making system aims to choose the most suitable option. According to the result of decision making and the driving styles of obstacle vehicles, motion planner will decide the optimal velocity and path within constraints for AV.

In the proposed integrated approach, the following assumptions are made:

a. In terms of decision making, only behavior planning is studied, and mission planning and route planning are not discussed in this work.
b. Only lane change scenario and straight lane scenarios are considered, curved roads will be discussed in the future.
c. The lane-change behaviors of obstacle vehicles are not included, thus only acceleration and deceleration behaviors are considered for obstacle vehicles.

Contributions of this paper are summarized as follows. Firstly, an integrated approach of decision making and motion planning is proposed for AVs. Stackelberg Game theory is used to solve the non-cooperative decision-making problem. Potential field model is applied in motion planning with MPC. Besides, different driving styles of obstacle vehicles are taken into consideration in the modeling process. Finally, the performance of the proposed approach is validated within different driving scenarios, demonstrating its feasibility, effectiveness and adaptivity.

## III. Development of Decision Making Based on Stackelberg Game Theory

Game theories have been widely used in decision making of multiple agents. For autonomous vehicles, there also exists the decision-making problem under complex traffic conditions. In this section, a non-cooperative game theory approach is investigated to solve the decision-making problem for AVs.

### A. Stackelberg Game Theory

When the vehicle ahead moves very slowly, the human driver of the host vehicle may choose to change lane. Before changing lanes, the human drivers will check the surrounding traffic environment. If the driver thinks it is safe for lane change, he or she will need to first interact with the surrounding vehicle through turning signal light. At this moment, the driver of the obstacle vehicle has two choices, i.e., acceleration or deceleration. If the driver of the obstacle vehicle is aggressive, he may choose acceleration. Then, the driver of the host vehicle should make decision: continuing the lane-change maneuver or staying on the current lane. Therefore, the decision making of the host vehicle need to take the aggressiveness of the obstacle vehicles into consideration.

The lane-change decision-making procedure mentioned above can be seen as a typical Stackelberg Game problem without cooperation. Next, we will model of the decision making for AVs by using the Stackelberg Game theory. Here, three key elements, namely, the player, action and cost, are considered in the Stackelberg Game model.

Targeting the lane change scenario of AVs, as shown in Fig. 2, the vehicle ahead is a bus which moves with a slow speed, and the host vehicle needs to make a decision, either following the front bus or change to the left lane. Formulating this scenario into a Stackelberg Game model, the host and obstacle vehicles can be seen as two players. The host vehicle is the leader, and the obstacle vehicle is the follower. The host vehicle can choose to change its lane or stay on the current lane. The obstacle vehicle can react by executing acceleration or deceleration. The action costs of the Stackelberg Game during decision making with the four possible interaction cases, which are related to the driving safety, ride comfort, travel efficiency and other factors, are listed in Table I.

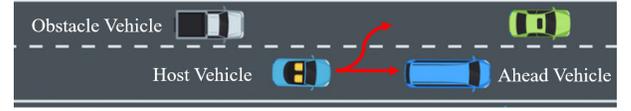

Fig. 2. Schematic diagram of the lane change decision-making for AVs.

TABLE I
ACTION COSTS WITH THE STACKELBERG GAME THEORY

| Action Cost | | Obstacle Vehicle | |
|---|---|---|---|
| | | Accelerate | Decelerate |
| Host Vehicle | Change Lane | (1, 6) | (3, 4) |
| | Stay | (2, 3) | (4, 5) |

In this game, both the host and obstacle vehicles make efforts to minimize their costs. For instance, if the host vehicle chooses to change its lane, and the obstacle vehicle reacts by accelerating, then the costs of the two vehicles will be 1 and 6, respectively. However, if the obstacle vehicle reacts by decelerating, then the costs will become 3 and 4, respectively. It can be found that the second choice is more suitable for the obstacle vehicle, since it results in a smaller cost. Similarly, if the host vehicle chooses to stay on its original lane, while the obstacle vehicle accelerates, then their costs will be 2 and 3, respectively. However, if the obstacle vehicle decelerates, then their costs will become 4 and 5, respectively. By contrast, the reaction of acceleration is more beneficial for the obstacle vehicle, due to the smaller cost. By comprehensive comparison of the listed four cases, in this scenario, the host vehicle will choose to stay on its current lane, and the obstacle vehicle will react by acceleration, in order to get gain smaller costs. Thus, the pair of action cost (2, 3) would be the solution of the Stackelberg equilibrium. It can be found that although the host vehicle has the priority to move first, its actual cost also relies on the correct reaction assessment of the obstacle vehicle.

### B. Cost Function of the Host Vehicle

In this paper, three factors, i.e., driving safety, ride comfort and travel efficiency, are considered to evaluate the decision-making cost of the vehicles.

For the host vehicle, the cost on driving safety consists of the longitudinal and lateral components, which is defined as

$$\Upsilon_{ds}^{hv} = ||\alpha|-1|\Upsilon_{ds-log}^{hv} + |\alpha|\Upsilon_{ds-lat}^{hv} \qquad (1)$$

where $\Upsilon_{ds-log}^{hv}$ and $\Upsilon_{ds-lat}^{hv}$ are the longitudinal and lateral safety cost, respectively. $\alpha$ is the decision making action of the host vehicle, $\alpha \in \{-1, 0, 1\} \coloneqq \{\text{change left, no lane change, change right}\}$.

The longitudinal cost of the driving safety for the host vehicle is a function of the longitudinal gap and relative velocity against the vehicle ahead, which can be represented by

$$\Upsilon_{ds-log}^{hv} = \kappa_{v-log}^{hv} \lambda_{\delta}^{hv} / [(\Delta v_{x,\delta}^{hv})^2 + \upsilon] + \kappa_{s-log}^{hv} / [(\Delta s_{x,\delta}^{hv})^2 + \upsilon] \qquad (2)$$

$$\Delta v_{x,\delta}^{hv} = v_{x,\delta}^{av} - v_{x,\delta}^{hv}$$
$$\Delta s_{x,\delta}^{hv} = X_{\delta}^{av} - X_{\delta}^{hv} - l_v \qquad (3)$$

$$\lambda_{\delta}^{hv} = \begin{cases} 1, & \Delta v_{x,\delta}^{hv} < 0 \\ 0, & \Delta v_{x,\delta}^{hv} \geq 0 \end{cases} \qquad (4)$$

where $v_{x,\delta}^{av}$ and $v_{x,\delta}^{hv}$ denote the longitudinal velocities of the front vehicle and host vehicle, respectively. $X_{\delta}^{av}$ and $X_{\delta}^{hv}$ are the longitudinal positions of the ahead vehicle and host vehicle, respectively. $\kappa_{v\text{-}log}^{hv}$ and $\kappa_{s-log}^{hv}$ are the weighting coefficients. $\upsilon$ is a small value set to avoid zero denominator in calculation. $l_v$ is a safety coefficient considering the length of the vehicle. $\delta$ denotes the current lane number of the host vehicle, where $\delta = 1, 2, 3$ denote the left lane, middle lane and right lane, respectively.

The lateral cost of the driving safety for the host vehicle is related to the longitudinal gap and relative velocity with respective to the obstacle vehicle, which can be expressed as

$$\Upsilon_{ds-lat}^{hv} = \kappa_{v\text{-}lat}^{hv} \lambda_{\delta+\alpha}^{hv} / [(\Delta v_{x,\delta+\alpha}^{hv})^2 + \upsilon] + \kappa_{s-lat}^{hv} / [(\Delta s_{x,\delta+\alpha}^{hv})^2 + \upsilon] \qquad (5)$$

$$\Delta v_{x,\delta+\alpha}^{hv} = v_{x,\delta}^{hv} - v_{x,\delta+\alpha}^{ov}$$
$$\Delta s_{x,\delta+\alpha}^{hv} = X_{\delta}^{hv} - X_{\delta+\alpha}^{ov} - l_v \qquad (6)$$

$$\lambda_{\delta+\alpha}^{hv} = \begin{cases} 1, & \Delta v_{x,\delta+\alpha}^{hv} < 0 \\ 0, & \Delta v_{x,\delta+\alpha}^{hv} \geq 0 \end{cases} \qquad (7)$$

where $v_{x,\delta+\alpha}^{ov}$ denotes the longitudinal velocity of the obstacle vehicle, $X_{\delta+\alpha}^{ov}$ denotes the longitudinal position of the obstacle vehicle, $\kappa_{v-lat}^{hv}$ and $\kappa_{s-lat}^{hv}$ are the weighting coefficients.

Besides, the lateral gap safety is not taken into consideration in the decision-making modeling and it will be discussed in motion planning.

The ride comfort cost of the host vehicle is related to the longitudinal acceleration and lateral acceleration directly, which is defined by

$$\Upsilon_{rc}^{hv} = \kappa_{a_x}^{hv}(a_{x,\delta}^{hv})^2 + |\alpha|\kappa_{a_y}^{hv}(a_{y,\delta}^{hv})^2 \qquad (8)$$

where $a_{x,\delta}^{hv}$ and $a_{y,\delta}^{hv}$ denote the longitudinal acceleration and lateral acceleration of the host vehicle. $\kappa_{a_x}^{hv}$ and $\kappa_{a_y}^{hv}$ are the weighting coefficients.

The travel efficiency cost of the host vehicle is related to the longitudinal velocity of the host vehicle, which is expressed as

$$\Upsilon_{pe}^{hv} = (v_{x,\delta}^{hv} - \overline{v}_{x,\delta}^{hv})^2 \qquad (9)$$

$$\overline{v}_{x,\delta}^{hv} = \min(v_{x,\delta}^{\max}, v_{x,\delta}^{av}) \qquad (10)$$

where $v_{x,\delta}^{\max}$ is the velocity limit on the lane $\delta$.

Finally, the cost function of the host vehicle is an integration of the costs on driving safety, ride comfort, and travel efficiency cost, which can be written as

$$\Upsilon^{hv} = \varpi_{ds}^{hv}\Upsilon_{ds}^{hv} + \varpi_{rc}^{hv}\Upsilon_{rc}^{hv} + \varpi_{pe}^{hv}\Upsilon_{pe}^{hv} \qquad (11)$$

where $\varpi_{ds}^{hv}$, $\varpi_{rc}^{hv}$, $\varpi_{pe}^{hv}$ are the weighting coefficients of driving safety, ride comfort, travel efficiency for the host vehicle.

### C. Cost Function of the Obstacle Vehicle

Similarly, the cost of the obstacle vehicle on driving safety includes the longitudinal and lateral components, as shown below:

$$\Upsilon_{ds}^{ov} = ||\alpha| - 1|\Upsilon_{ds-log}^{ov} + |\alpha|\Upsilon_{ds-lat}^{ov} \qquad (12)$$

where $\Upsilon_{ds-log}^{ov}$ and $\Upsilon_{ds-lat}^{ov}$ are the costs of the longitudinal and lateral driving safety, respectively.

The longitudinal component can be written as

$$\Upsilon_{ds-log}^{ov} = \kappa_{v\text{-}log}^{ov} \lambda_{\delta+\alpha}^{ov} / [(\Delta v_{x,\delta+\alpha}^{ov})^2 + \upsilon] + \kappa_{s-log}^{ov} / [(\Delta s_{x,\delta+\alpha}^{ov})^2 + \upsilon] \qquad (13)$$

$$\Delta v_{x,\delta+\alpha}^{ov} = v_{x,\delta+\alpha}^{ao} - v_{x,\delta+\alpha}^{ov}$$
$$\Delta s_{x,\delta+\alpha}^{ov} = X_{\delta+\alpha}^{ao} - X_{\delta+\alpha}^{ov} - l_v \qquad (14)$$

$$\lambda_{\delta+\alpha}^{ov} = \begin{cases} 1, & \Delta v_{x,\delta+\alpha}^{ov} < 0 \\ 0, & \Delta v_{x,\delta+\alpha}^{ov} \geq 0 \end{cases} \qquad (15)$$

where $v_{x,\delta+\alpha}^{ov}$ and $v_{x,\delta+\alpha}^{ao}$ denote the longitudinal velocities of the obstacle vehicle and its front vehicle, respectively. $X_{\delta+\alpha}^{ov}$ and $X_{\delta+\alpha}^{ao}$ denote the longitudinal positions of the obstacle vehicle and its vehicle ahead, respectively. $\kappa_{v\text{-}log}^{ov}$ and $\kappa_{s-log}^{ov}$ denote the weighting coefficients.

Assuming that the obstacle vehicle does not change lanes and only takes action of acceleration or deceleration, the lateral cost of the obstacle vehicle on driving safety is equivalent to the that of the host vehicle.

$$\Upsilon_{ds-lat}^{ov} = \Upsilon_{ds-lat}^{hv} \qquad (16)$$

The cost of the obstacle vehicle on ride comfort only correlates to the longitudinal motion, which can be expressed as

$$\Upsilon_{rc}^{ov} = \kappa_{a_x}^{ov}(a_{x,\delta+\alpha}^{ov})^2 \qquad (17)$$

where $a_{x,\delta+\alpha}^{ov}$ is the longitudinal acceleration of the host vehicle, and $\kappa_{a_x}^{ov}$ is the weighting coefficient.

Similarly, the cost of the obstacle vehicle on travel efficiency can be given by

$$\Upsilon_{pe}^{ov} = (v_{x,\delta+\alpha}^{ov} - \overline{v}_{x,\delta+\alpha}^{ov})^2 \qquad (18)$$

$$\overline{v}_{x,\delta+\alpha}^{ov} = \min(v_{x,\delta+\alpha}^{\max}, v_{x,\delta+\alpha}^{ao}) \qquad (19)$$

where $v_{x,\delta+\alpha}^{\max}$ is the velocity limit on the lane $\delta+\alpha$.

Furthermore, the cost function of the obstacle vehicle is an integrated consideration of the driving safety, ride comfort and travel efficiency, which can be expressed as

$$\Upsilon^{ov} = \varpi_{ds}^{ov}\Upsilon_{ds}^{ov} + \varpi_{rc}^{ov}\Upsilon_{rc}^{ov} + \varpi_{pe}^{ov}\Upsilon_{pe}^{ov} \quad (20)$$

where $\varpi_{ds}^{ov}$, $\varpi_{rc}^{ov}$, $\varpi_{pe}^{ov}$ are the weighting coefficients of driving safety, ride comfort, travel efficiency for the obstacle vehicle.

### D. Decision Making using Stackelberg Game Theory

In the lane-change decision making process, one obstacle vehicle is taken into account firstly. Then, the host vehicle and the obstacle vehicle are formulated into a 2-player Stackelberg Game, which is a bilevel optimization problem. It can be expressed as follows.

$$(a_{x,\delta}^{hv*}, \alpha^*) = \arg\min_{a_{x,\delta}^{hv}, \alpha}(\max_{a_{x,\delta+\alpha}^{ov} \in \gamma^2(a_{x,\delta}^{hv}, \alpha)} \Upsilon^{hv}(a_{x,\delta}^{hv}, \alpha, a_{x,\delta+\alpha}^{ov})) \quad (21)$$

$$\begin{cases} \gamma^2(a_{x,\delta}^{hv}, \alpha) \square \{\zeta \in \Phi^2 : \Upsilon^{ov}(a_{x,\delta}^{hv}, \alpha, \zeta) \leq \Upsilon^{ov}(a_{x,\delta}^{hv}, \alpha, a_{x,\delta+\alpha}^{ov}), \\ \quad \forall a_{x,\delta+\alpha}^{ov} \in \Phi^2 \} \\ (\alpha+1)\alpha(\alpha-1) = 0 \end{cases} \quad (22)$$

where $a_{x,\delta}^{hv*}$ is the optimal longitudinal acceleration of the host vehicle, $\alpha^*$ is the optimal lane change decision of the host vehicle, $\gamma^2(a_{x,\delta}^{hv}, \alpha)$ denotes the optimal action selection of the obstacle vehicle given the decision made by the host vehicle, and $\Phi^2$ denotes the action selections of the obstacle vehicle.

Additionally, constraints on velocity and acceleration should be taken into consideration.

$$v_{x,\delta} \in [0, \ v_{x,\delta}^{max}], \ a_{x,\delta} \in [a_{x,\delta}^{min}, \ a_{x,\delta}^{max}] \quad (23)$$

where $a_{x,\delta}^{min}$ and $a_{x,\delta}^{max}$ denote the lower and upper boundaries of the acceleration, respectively.

If it is a two-lane road or the host vehicle moves in the most marginal lane, there should be only one option to change the lane, i.e. either right or left lane change. And only one obstacle vehicle needs to be considered. However, if the host vehicle moves on the middle lane and the total number of the lanes is more than two, then the host vehicle can change to either the left or right lane, and two obstacle vehicles should be taken into consideration. The optimization problem can be expressed as:

$$(a_{x,\delta}^{hv*}, \alpha^*) = \arg\min_{a_{x,\delta}^{hv}, \alpha}(\max_{a_{x,\delta+\alpha}^{i} \in \gamma^2(a_{x,\delta}^{hv}, \alpha)} \Upsilon^{hv}(a_{x,\delta}^{hv}, \alpha, a_{x,\delta+\alpha}^{i})) \quad (24)$$

$$\begin{cases} \gamma^2(a_{x,\delta}^{hv}, \alpha) = \{\zeta \in \Phi^2 : \Upsilon^i(a_{x,\delta}^{hv}, \alpha, \zeta) \leq \Upsilon^i(a_{x,\delta}^{hv}, \alpha, a_{x,\delta+\alpha}^{i}), \\ \quad \forall a_{x,\delta+\alpha}^i \in \Phi^2\}, \ (i = lov, rov) \\ (\alpha+1)\alpha(\alpha-1) = 0 \end{cases} \quad (25)$$

where *lov* and *rov* denote the left obstacle vehicle and right obstacle vehicle, respectively.

It can be found that aforementioned optimization-based decision making problem is highly coupled with motion planning of the vehicles. Therefore, the lane-change decision making should be combined with motion planning together. And this will be discussed in Section V in details.

### IV. MOTION PLANNING BASED ON POTENTIAL FIELD MODEL

Potential field is an effective method to model the dynamics of the vehicle and describe its interaction with surrounding obstacles. In this section, the potential field model is combined with MPC for the optimal path planning of AVs.

### A. The Potential Field Model

The potential field model of obstacle vehicle is defined as follows [34].

$$P^{ov}(X,Y) = a^{ov}e^{\eta} \quad (26)$$

$$\eta = -\left\{\frac{(X-X^{ov})^2}{2\rho_X^2} + \frac{(Y-Y^{ov})^2}{2\rho_Y^2}\right\}^b + cv_x^{ov}\xi \quad (27)$$

$$\xi = k^{ov}\frac{(X-X^{ov})^2}{2\rho_X^2} / \sqrt{\frac{(X-X^{ov})^2}{2\rho_X^2} + \frac{(Y-Y^{ov})^2}{2\rho_Y^2}} \quad (28)$$

$$k^{ov} = \begin{cases} -1, & (X-X^{ov}) < 0 \\ 1, & (X-X^{ov}) \geq 0 \end{cases} \quad (29)$$

where $P^{ov}(X,Y)$ denotes the potential field value of the obstacle vehicle at the position $(X, Y)$ in the geodetic coordinate system. $(X^{ov}, Y^{ov})$ denotes the CoG (center of gravity) position of the obstacle vehicle. $a^{ov}$ is the maximum potential field value of the obstacle vehicle. $\rho_X$ and $\rho_Y$ are the convergence coefficients along the directions of $X$ and $Y$, respectively. $v_x^{ov}$ is the longitudinal velocity of the obstacle vehicle. $b$ is the shape coefficient.

According to Eq. (26) to Eq. (29), the 3D map of the potential field model for obstacle vehicle is illustrated in Fig. 3.

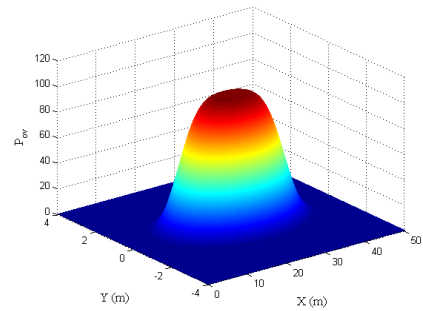

Fig. 3. An example of the 3D map of the potential field model for the obstacle vehicle.

Additionally, the potential field model of the road can be defined as

$$P^r(X,Y) = a^r e^{(-d+d^r+0.5W)} \quad (30)$$

where $a^r$ is the maximum value of the potential field of the road, $d$ is the minimum distance from the position $(X, Y)$ to the lane mark, $d^r$ is the safety threshold, and $W$ is the width of the vehicle.

Based on the Eq. (30), the 3D map of the potential field model for the three-lane road is presented in Fig. 4.

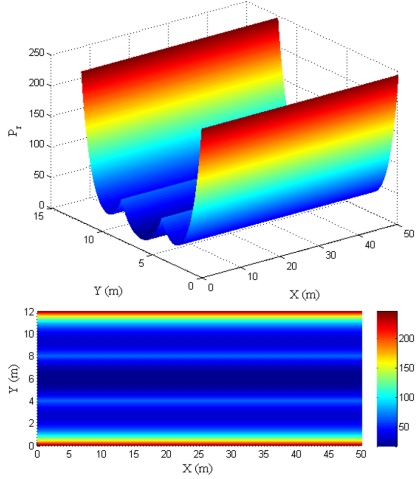

Fig. 4. 3D map of the potential field model for road with three lanes.

Combining the potential field models of all the obstacles and roads, it yields an integrated function:

$$P^c(X,Y) = \sum_{i=1}^{m} P_i^{ov}(X,Y) + \sum_{j=1}^{n} P_j^r(X,Y) \qquad (31)$$

where $m$ is the number of the obstacle vehicles, and $n$ is the number of lane lines.

Fig. 5 shows the potential filed model of three vehicles on a three-lane highway. It can be seen that the value of the potential field is very large near the obstacles and road boundaries. While the value of the potential field for the free space is very small.

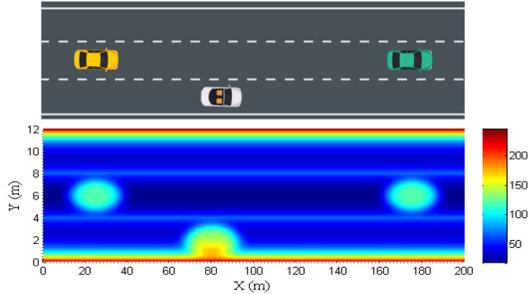

Fig. 5. Potential field model of three vehicles on a high way road with three lanes.

### B. The Motion Planning Model

As mentioned before, for AVs, motion planning consists of longitudinal motion planning and lateral path planning. The longitudinal motion planned for AVs is the acceleration or velocity. For decision making, in Section III, the longitudinal acceleration of the host vehicle has been considered and needs to be solved. That is to say, the longitudinal motion planning has been coupled with the decision making module.

For design of the motion planning, the following simplified kinematic model is proposed.

$$\begin{cases} \dot{v}_{x,\delta}^{hv} = a_{x,\delta}^{hv} \\ \dot{\varphi}_{\delta}^{hv} = a_{y,\delta}^{hv} / v_{x,\delta}^{hv} \\ \dot{X}_{\delta}^{hv} = v_{x,\delta}^{hv} \cos \varphi_{\delta}^{hv} \\ \dot{Y}_{\delta}^{hv} = v_{x,\delta}^{hv} \sin \varphi_{\delta}^{hv} \end{cases} \qquad (32)$$

where $\varphi_{\delta}^{hv}$ is the heading angle of the host vehicle.

### C. Prediction of the Planned Path

To predict the planned path, Eq. (32) is rewritten as

$$\dot{x}(t) = f[x(t), u(t)] \qquad (33)$$

$$f[x(t),u(t)] = \begin{bmatrix} a_{x,\delta}^{hv} \\ a_{y,\delta}^{hv} / v_{x,\delta}^{hv} \\ v_{x,\delta}^{hv} \cos \varphi_{\delta}^{hv} \\ v_{x,\delta}^{hv} \sin \varphi_{\delta}^{hv} \end{bmatrix} \qquad (34)$$

where the state vector $x = \begin{bmatrix} v_{x,\delta}^{hv} & \varphi_{\delta}^{hv} & X_{\delta}^{hv} & Y_{\delta}^{hv} \end{bmatrix}^T$, and the control vector $u = a_{y,\delta}^{hv}$.

To get the discretized system, Eq. (33) can be expressed as

$$x(k+1) = x(k) + \Delta T \cdot f[x(k), u(k)] \qquad (35)$$

where $\Delta T$ is the sampling time.

Besides, the output vector $y$ is the integrated potential field value associated with position coordinates.

$$y(k) = g[x(k), u(k)] = P^c(X_{\delta}^{hv}(k), Y_{\delta}^{hv}(k)) \qquad (36)$$

Then, MPC is used to predict the state with multiple steps. The prediction outputs at the time step $k$ are described as

$$\begin{aligned} y(k+1|k) &= g[x(k+1|k), u(k|k)] \\ y(k+2|k) &= g[x(k+2|k), u(k+1|k)] \\ &\vdots \qquad\qquad \vdots \\ y(k+N_c|k) &= g[x(k+N_c|k), u(k+N_c-1|k)] \\ y(k+N_c+1|k) &= g[x(k+N_c+1|k), u(k+N_c-1|k)] \\ &\vdots \qquad\qquad \vdots \\ y(k+N_p|k) &= g[x(k+N_p|k), u(k+N_c-1|k)] \end{aligned} \qquad (37)$$

Moreover, the output sequence and control sequence are defined by

$$\mathbf{y}(k) = [y(k+1|k), \; y(k+2|k), \; \cdots, \; y(k+N_p|k)]^T \qquad (38)$$

$$\mathbf{u}(k) = [u(k|k), \; u(k+1|k), \; \cdots, \; u(k+N_c-1|k)]^T \qquad (39)$$

where $N_p$ and $N_c$ are the prediction horizon and control horizon, respectively. $N_p > N_c$.

At the time step $k$, taking the control sequence into account, the cost function for motion planning is defined.

$$J^p(k) = \mathbf{y}^T(k)Q_1\mathbf{y}(k) + \Delta\mathbf{Y}^T(k)Q_2\Delta\mathbf{Y}(k) + \mathbf{u}^T(k)R\mathbf{u}(k) \quad (40)$$

where $Q_1$ and $Q_2$ are the output weighting matrices, $R$ is the control input variation weighting matrix, $\Delta\mathbf{Y}$ is the lateral distance error sequence between the planned path and the center line of the lane.

Besides, the motion planning can be transformed into a constrained optimization problem.

$$\mathbf{u}(k) = \arg\min_{\mathbf{u}(k)} J^p(k) \quad (41)$$

subject to

$$x(k+i\,|\,k) = A_k x(k+i-1\,|\,k) + B_k u(k+i-1\,|\,k)$$
$$y(k+i-1\,|\,k) = g\left[x(k+i-1\,|\,k), u(k+i-1\,|\,k)\right]$$
$$u_{\min} \leq u(k+i-1\,|\,k) \leq u_{\max}$$

where $u_{\min}$ and $u_{\max}$ denote the minimum and maximum values of the control vector.

After solving the above constrained optimization problem, the following optimal control sequence can be achieved.

$$\mathbf{u}^*(k) = \left[u^*(k\,|\,k),\ u^*(k+1\,|\,k),\ \cdots,\ u^*(k+N_c-1\,|\,k)\right]^T \quad (42)$$

Then, it is used to predict the planned path. At the next time step $k+1$, a new optimization is started over a shifted prediction horizon again with the updated state $x(k+1\,|\,k+1)$.

## V. INTEGRATED SOLUTION CONSIDERING DISTINGUISHED SOCIAL BEHAVIORS

Social behaviors, which can be reflected by different driving styles of obstacle vehicles, would affect the decision making and planning of AVs. Therefore, it would be good to characterize driving styles and embed their key features into the integrated decision making and motion planning algorithm.

### A. Decision Making Considering Different Social Behaviors of Obstacle Vehicles

As mentioned in Section II, three different driving styles, i.e., aggressive, normal and cautious, indicating distinguished social behaviors, are defined. In the decision-making problem, travel efficiency is defined as the dominant objective for aggressive drivers. Cautious drivers are defined to concern more about safety. Normal drivers are regarded as positioning in between.

In the modeling process of decision making, the driving style of the obstacle vehicle is associated with the driving safety, ride comfort and travel efficiency by using different settings of weighting coefficients $\varpi_{ds}^{ov}$, $\varpi_{rc}^{ov}$, $\varpi_{pe}^{ov}$. Reference to [33], the weighting selections of the three driving styles are listed in TABLE II.

TABLE II
WEIGHTING COEFFICIENTS OF DIFFERENT DRIVING STYLES

| Weighting Coefficients | Aggressive | Normal | Cautious |
|---|---|---|---|
| $\varpi_{ds}^{ov}$ | 10% | 50% | 70% |
| $\varpi_{rc}^{ov}$ | 10% | 30% | 20% |
| $\varpi_{pe}^{ov}$ | 80% | 20% | 10% |

### B. Motion Planning Considering Different Social Behaviors of Obstacle Vehicles

In the modeling process of motion planning, driving styles of obstacle vehicles are associated to the potential field models. Fig. 6 shows the three potential filed models, representing different driving styles of obstacle vehicles with the same velocity. For aggressive driving style, it has the widest distribution along the vehicle's moving direction, indicating that it would be more dangerous when approaching the aggressive vehicle. In contrast, the cautious style has the smallest distribution within the potential field. It means that it would be much safer when interacting with cautious drivers, compared to aggressive ones.

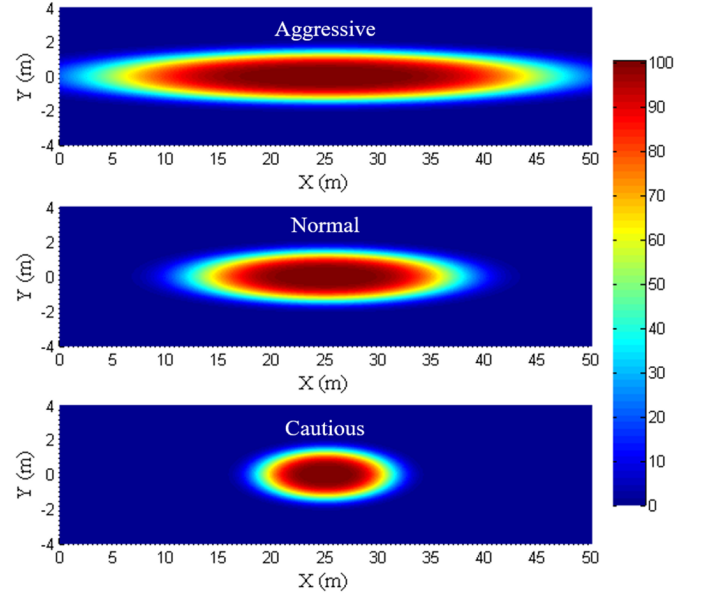

Fig. 6. Potential field models with different driving styles.

### C. Integration Decision Making and Path Planning and Its Solution

**Stackelberg Game Theoretic Optimization for Decision Making**

$$(a_{x,\delta}^{hv*}(k), \alpha^*(k)) = \arg\min_{a_{x,\delta}^{hv}, \alpha}(\max_{a_{x,\delta}^{i} \in \gamma^2(a_{x,\delta}^{hv}, \alpha)} \Upsilon^{hv}(a_{x,\delta}^{hv}(k), \alpha(k), a_{x,\delta}^{i}(k)))$$

s.t.
$$\gamma^2(a_{x,\delta}^{hv}(k), \alpha(k)) \square \{\zeta \in \Phi^2 : \Upsilon^{ov}(a_{x,\delta}^{hv}(k), \alpha(k), \zeta(k))$$
$$\leq \Upsilon^{ov}(a_{x,\delta}^{hv}(k), \alpha(k), a_{x,\delta}^{ov}(k)), \forall a_{x,\delta}^{ov}(k) \in \Phi^2\}$$
$$(\alpha(k)+1)\alpha(k)(\alpha(k)-1) = 0 \quad v_{x,\delta} \in [0,\ v_{x,\delta}^{\max}]\quad a_{x,\delta} \in [a_{x,\delta}^{\min},\ a_{x,\delta}^{\max}]$$

$\alpha^*$, $a_{x,\delta}^{hv*}$ ↓   ↑ $a_{y,\delta}^{hv*}$, $v_{x,\delta}^{hv}$, $X_{\delta}^{hv}$, $Y_{\delta}^{hv}$

**MPC-Based Optimization for Motion Planning**

$$\mathbf{u}(k) = \arg\min_{\mathbf{u}(k)} J^p(k)$$

s.t.
$$x(k+i\,|\,k) = A_k x(k+i-1\,|\,k) + B_k u(k+i-1\,|\,k)$$
$$y(k+i-1\,|\,k) = g\left[x(k+i-1\,|\,k), u(k+i-1\,|\,k)\right]$$
$$u_{\min} \leq u(k+i-1\,|\,k) \leq u_{\max}$$

Fig. 7. Integration of decision making and path planning.

As shown in Fig. 7, now decision making and motion planning modules are highly coupled with consideration of the social behaviors. The solution to the decision-making problem $\alpha^*$ and $a_{x,\delta}^{hv*}$ are the inputs of motion-planning system. In the meantime, the generated states of motion-planning module $a_{y,\delta}^{hv*}$, $v_{x,\delta}^{hv}$, $X_{\delta}^{hv}$ and $Y_{\delta}^{hv}$ are the feedback states to decision-making system.

Therefore, the entire computation process is a closed-loop iterative optimization process within multi-constraints. In this study, this formulated optimization problem is solved using the efficient evolutionary algorithm [35].

## VI. TESTING AND DISCUSSION

The feasibility and effectiveness of the integrated approach of decision making and path planning for AVs are validated in this section through different testing scenarios. As shown in Fig. 8, three different testing scenarios are illustrated. Case 1 and Case 2 are the scenarios of a two-lane highway, and Case 3 is a three-lane one. These three testing cases are typical conditions that are usually adopted to test the capability and performance of decision making and path planning. From Case 1 to Case 3, the complexity of the scenario increases, which is also able to test the adaptivity of the developed algorithms. All the testing cases are conducted on the platform of Matlab-Simulink. In each case, different social behaviors of obstacle vehicles will be discussed.

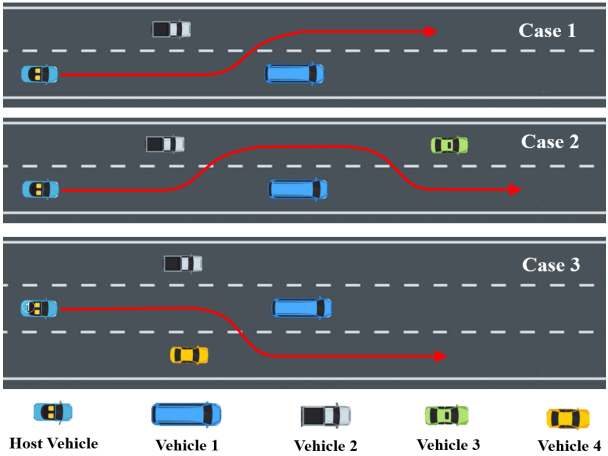

Fig. 8. Three test cases for decision making and motion planning.

### A. Testing Case 1

In this case, a common seen single lane-change scenario is studied. At the initial moment, the host vehicle (HV) and vehicle 1 (V1) move on the same lane with instantaneous velocities of 20 m/s and 15 m/s, respectively. The initial gap between the two vehicles is 50 m. Vehicle 2 (V2) moves on the left adjacent lane with an instantaneous velocity of 12 m/s, and the initial gap between HV and V2 is 2 m (HV is in front of V2). Correspondingly, V1 is the vehicle ahead, and vehicle 2 is set as the obstacle vehicle. Since the vehicle ahead moves slower, HV must make a decision to decelerate and follow the vehicle ahead, or changing a lane and overtake. The decision making is also affected by the reaction behavior of the obstacle vehicle. In this case, V1 is assumed to move forward with a constant speed. Testing results are illustrated in Figs. 9 and 10 in details.

The obstacle V2 with three different driving styles would lead to different result of decision-making and motion-planning of HV. It can be seen from Figs. 9 and 10 that if V2 is aggressive, then V2 would chooses a sudden acceleration action to stop the lane-change intention of HV. As a result, HV has to follow the vehicle ahead V1. After 6 s, HV needs to slow down to keep a safe distance against V1. If V2's driving style is set as normal, it will not speed up with a large acceleration. Then, the HV has enough safe space to do lane change. While if the driving style of V2 becomes cautious, then the HV has an even larger space to change lane. And its initial time of executing lane-change is earlier compared to the second scenario, since the acceleration of obstacle V2 is smaller.

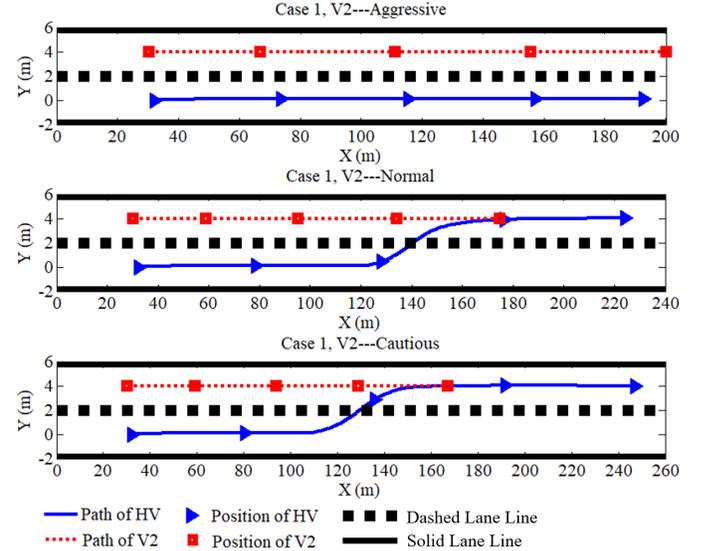

Fig. 9. Results of decision making and path planning considering different driving styles of obstacle vehicles in Case 1.

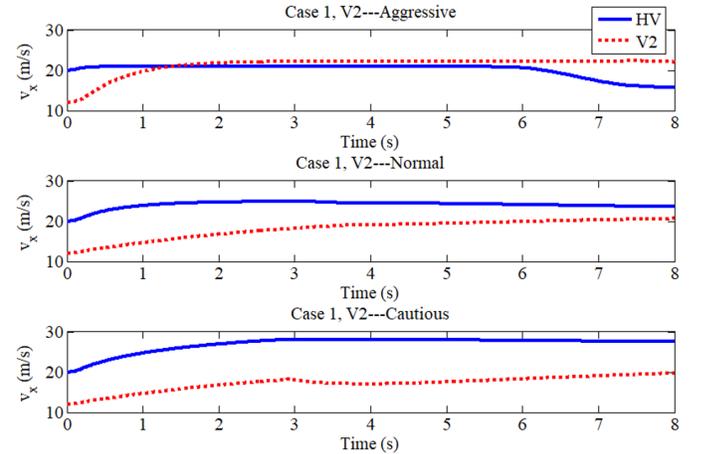

Fig. 10. Testing results of decision making and planning considering different driving styles of obstacle vehicles in Case 1.

### B. Testing Case 2

Case 2 is a double lane-change scenario. It can be seen as a supplementary to Case 1. Suppose that the V2 is not aggressive, then HV has changed lanes successfully. However, another obstacle vehicle (V3) moves in front of HV, which means the V3 becomes the vehicle ahead after HV finishing the initial lane change. Since V3 moves slower, the HV needs to make another decision, decelerating and following the vehicle ahead, or changing to the right lane and overtaking. At this moment, V1 becomes the obstacle vehicle, and the decision making should take the behavior of the V1 into consideration. In this case, V3 is assumed to move forward with a constant velocity of 15 m/s. The initial gap between V2 and V3 is 105 m. Tests results in this case are illustrated in Figs. 11 and 12.

In this case, we also simulate the scenarios of the V1 with three different driving styles. Although V1 moves forward at a constant speed at first, it will have reactive behaviors if it finds HV has a lane change intention. When V1's driving style is set as aggressive, it will accelerate to prevent HV's lane-change behavior. Hence, HV has to move on its current lane and give up lane change. Once HV approaching the vehicle ahead, it has to decelerate to keep a safe distance against it. If the driving style of V1 is normal, it will not fiercely fight for the right of way against others. As a result, it will give way to the HV. We can see from Fig. 11 that HV finishes its double lane-change action at the position of 320 m. For V1 with a cautious style, HV would spend much less time and shorter distance to complete the double lane-change behavior due to V1's conservative social behavior.

## C. Testing Case 3

In this case, a more complicated three-lane highway scenario is taken into account. At the initial time, HV and V1 move on the middle lane with instantaneous velocities of 20 m/s and 15 m/s, respectively. The initial gap between the two vehicles is 30 m. On the left adjacent lane, V2 moves with an instantaneous velocity of 12 m/s, and the initial gap between HV and V2 is 2 m (the HV is in front of V2). On the right adjacent lane, vehicle 4 (V4) moves with an instantaneous velocity 13 m/s, and the initial gap between HV and V4 is 3 m (the V4 is in front of HV). In this case, V1 is the ahead vehicle, both V2 and V4 are obstacle vehicles. Considering that the velocity of ahead vehicle is smaller than that of the HV, HV needs to make decisions. Both the behaviors of V2 and V4 must be considered within the decision-making process of the HV. Figs. 13 and 14 show the detailed testing results.

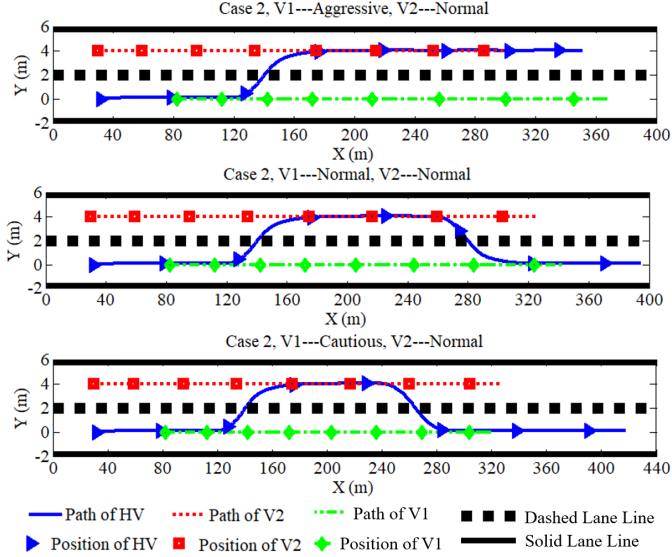

Fig. 11. Results of decision making and path planning considering different driving styles of obstacle vehicles in Case 2.

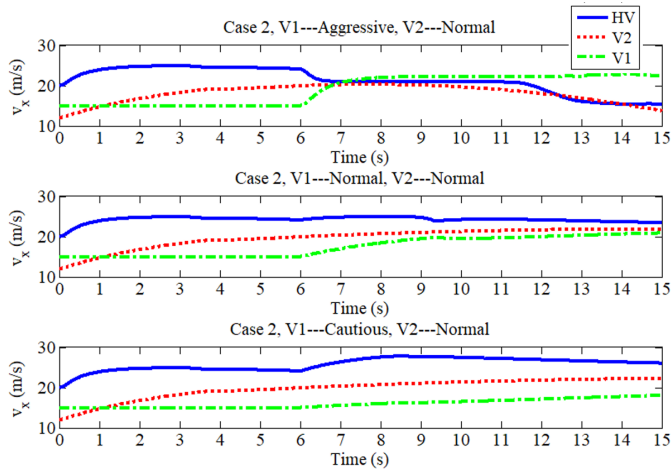

Fig. 12. Testing results of decision making and planning considering different driving styles of obstacle vehicles in Case 2.

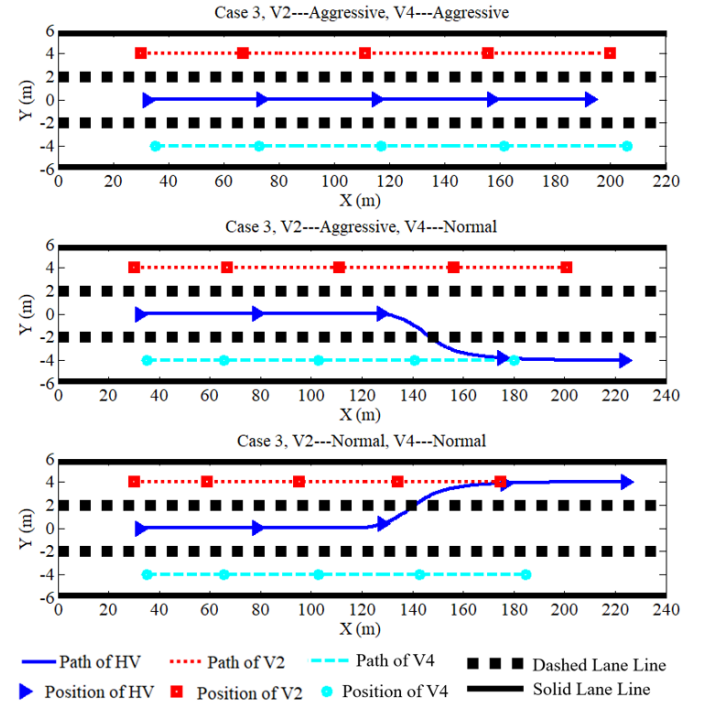

Fig. 13. Results of decision making and path planning considering different driving styles of obstacle vehicles in Case 3.

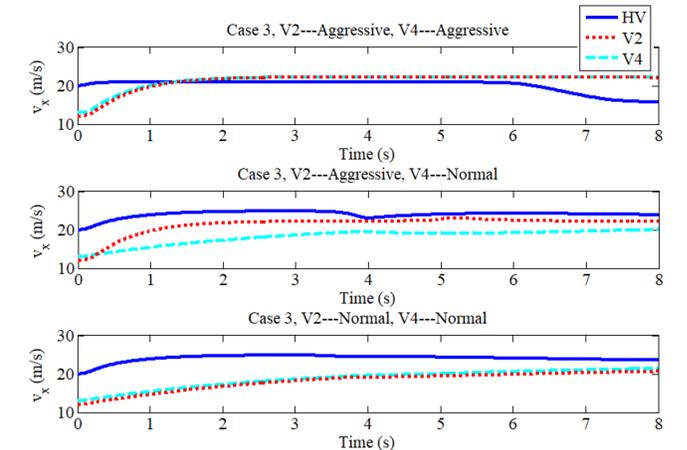

Fig. 14. Testing results of decision making and planning considering different driving styles of obstacle vehicles in Case 3.

The first scenario is set that the driving styles of both V2 and V4 are aggressive, such that they would not give way to the HV. As a result, HV has to follow the front vehicle and slow down when needed to keep a safe distance. The second scenario is the driving style of V2 is considered as aggressive, while V4 is set as normal. In this condition, V4 would give up acceleration and give way to the HV. The HV can change lanes successfully with enough safe space. The third scenario is that both V2 and V4 are set to normal styles. Based on the testing results shown in Fig. 14, the HV would choose left-side lane change due to the smaller cost generated on the left lane.

## VII. Conclusions and Future Work

This paper presents an integrated approach of decision making and path planning for AVs. Social behaviors, which are reflected by three different driving styles of obstacle vehicles, are defined. Considering the social behaviors of vehicles, Stackelberg Game model is established to design the decision-making algorithm for AVs. The potential field model is adopted to describe the social characterizations of vehicles and embedded in the motion-planning module. And MPC is used the speed and path prediction of the AV. Finally, the decision making and motion planning are integrated and transformed into a closed-loop interative optimization problem with multi-constraints. Testing is carried in three different cases in order to evaluate the performance of the proposed integrated approach. Testing results show that the developed method is able to deal with reasonable decision making and safe path planning for AVs under various social behaviors of surrounding traffic participants, validating the feasibility and effectiveness of the proposed approach.

Our future work will focus on the improvement of the proposed approach with the consideration of more complexed driving conditions, to further improve the capability of decision making for connected autonomous vehicles.